# Investigation of indoor localization with ambient FM radio stations

Andrei Popleteev, Venet Osmani, Oscar Mayora
Ubiquitous Interactions group
CREATE-NET
Trento, Italy
firstname.lastname@create-net.org

*Abstract*— **Localization plays an essential role in many ubiquitous computing applications. While the outdoor location-aware services based on GPS are becoming increasingly popular, their proliferation to indoor environments is limited due to the lack of widely available indoor localization systems. The de-facto standard for indoor positioning is based on Wi-Fi and while other localization alternatives exist, they either require expensive hardware or provide a low accuracy. This paper presents an investigation into localization system that leverages signals of broadcasting FM radio stations. The FM stations provide a worldwide coverage, while FM tuners are readily available in many mobile devices. The experimental results show that FM radio can be used for indoor localization, while providing longer battery life than Wi-Fi, making FM an alternative to consider for positioning.**

## I. Introduction

Localization is an important area in ubiquitous computing, having been rooted in the early days of context-awareness research. One of the first definitions of context [1] refers to location as one of the four pillars of context in addition to time, activity and identity. Despite the relatively long tradition, localization remains an active area of research [2]. The focus of this research has been increasingly shifting towards indoor localization. This is because the wide adoption of the Global Positioning System (GPS) in mobile devices, combined with Wi-Fi and cellular networks have, in large part, solved the problem of outdoor localization. However, numerous ubiquitous computing applications rely on indoor positioning, where the GPS signal is too weak.

Despite the substantial research and development efforts, the existing indoor positioning systems remain unsuitable for wide adoption. Wi-Fi-based positioning is the de-facto standard for indoor localization. However, Wi-Fi-based systems have a number of issues, including high power consumption, limited coverage, and are prone to interference [3]. Other indoor positioning alternatives, based on RFID, infrared, ultrasound or ultra-wide band (UWB), require specialized hardware and dedicated infrastructure, resulting in high costs for wide adoption. The positioning systems based on cellular networks, in turn, provide a good coverage and can use the existing infrastructure, but suffer from low positioning accuracy.

Considering these issues, we describe and evaluate an alternative approach for indoor localization based on frequency modulated (FM) radio. Our localization system relies on the existing infrastructure of broadcasting FM stations as signal sources and embedded FM radio modules on client devices.

Current research literature shows that FM positioning using broadcasting stations has been investigated in outdoors context only [4, 15]. The low accuracy achieved outdoors discouraged investigation of FM for indoor localization. This may also explain the low number of publications in this area and no reported work on the use of FM for indoor positioning.

However, the results of outdoor localization cannot be directly projected onto indoor scenarios, as these environments are notably different with regard to signal propagation [5]. Furthermore, the performance of FM radio for indoor localization cannot be predicted from other technologies, such as Wi-Fi or GSM. This is because FM radio uses significantly lower frequencies (100 MHz) than GSM (0.9 GHz) or Wi-Fi (2.4 GHz) leading to different signal propagation and resulting in the following properties:

i) FM radio signals are less affected by weather conditions, such as rain or fog, in comparison to Wi-Fi or GSM [6, 8]; ii) low-frequency radio waves are less sensitive to terrain conditions, such as woodland and tree foliage [7] while, GSM and Wi-Fi signal propagation can be affected even by the movement of leaves [7]; iii) amount of attenuation of radio waves, caused by building materials is directly proportional to the operating frequency [10] therefore, FM signals penetrate walls more easily in comparison to Wi-Fi or GSM. This ensures high availability of FM positioning signals in indoor settings; iv) the FM wavelength of around 3 m (from 2.78 m to 3.43 m in Europe and US) interacts differently with most indoor objects in comparison to the wavelength of 0.12 m of Wi-Fi waves. At low frequencies, when the obstacles are small compared to the wavelength, they do not interact significantly with the electromagnetic fields of the wave [10]. However, when the size of an obstacle is close to the wavelength, interaction becomes strong, resulting in complex interference patterns

[10]. This means that majority of small indoor objects are transparent to long FM radio waves, but interact and cause interference with shorter Wi-Fi and GSM waves; v) overall, FM radio is a popular and well-established technology. Broadcasting FM stations provide almost ubiquitous coverage, in populated areas worldwide. FM receivers are already embedded in many mobile devices, they have low power consumption and do not interfere with sensitive equipment or other wireless technologies. These properties make FM radio an interesting technology to investigate for a positioning system.

In the rest of this paper, we provide an overview of the indoor positioning systems based on Wi-Fi and cellular network signals, as well as the existing FM-based localization systems. Then we detail our positioning approach based on broadcasting FM stations. We evaluate its performance on different scale environments and at different time frames and provide a comparison to Wi-Fi and GSM systems. Finally we analyze the impact of FM, Wi-Fi and GSM localization modes on mobile device's battery life.

## II. RELATED WORK

### A. Wi-Fi

IEEE 802.11 standard for wireless networks is a popular basis for indoor positioning systems. Their popularity among the research community can be explained by frequently available network infrastructure, Wi-Fi enabled mobile devices, good localization performance and their use in ubiquitous computing applications. Wi-Fi networks are deployed in many office buildings and homes, and positioning systems can exploit the already available beacons.

The RADAR project [9] pioneered Wi-Fi positioning. This work used received signal strength (RSS) fingerprinting approach and achieved 2.94 m median error [9]. Ferris et al. [12] designed a Wi-Fi localization system using Gaussian processes in conjunction with graph-based tracking. They modeled users moving through the rooms on the same floor, as well as more complicated patterns of moving, such as going up and downstairs. The average error of test data was 2.12 meters. Work in [13] used advanced probabilistic methods and Wi-Fi based positioning reached an accuracy of 1.2 m.

Brunato and Battiti [14] compared the performance of Wi-Fi fingerprinting localization using several machine learning methods, such as multi-layer perceptron (MLP), support vector machine (SVM) and k-nearest neighbor (kNN). The SVM approach demonstrated the best median accuracy of 2.75m. Notably, the median performance of a simple unweighted kNN classifier was only 0.16 m less than the SVM's result; the 95th percentile errors were equal.

While the accuracy degradation is typical for fingerprinting based systems, the impact of each factor varies, and is dependent on signal frequency: when the obstacles are small in comparison to wavelength, their interaction with the wave is negligible [10] and vice versa. Therefore, environmental factors have smaller impact on lower-frequency FM radio waves. However, most indoor radio wave propagation measurements have been done for frequencies above 1 GHz [5] and there is a lack of experimental results for lower frequencies.

### B. Cellular networks

In comparison to Wi-Fi, cellular networks, such as GSM and CDMA, provide significantly better coverage. However, for a long time they were not considered for indoor localization due to the low accuracy achieved in outdoor settings [17, 19, 22].

The first results for indoor GSM positioning performance have been published in [17]. The proposed approach employed so-called wide fingerprints, which include the RSSI readings from 6 strongest stations, extended (widened) by RSSI data from up to 35 weaker GSM channels. The experimental results for different buildings have demonstrated a median accuracy from 3.4 m to 11 m with six strongest stations, and from 2.5 m to 5.4 m with wide fingerprints. In 3 out of 7 tests the GSM accuracy with wide fingerprints was better than the Wi-Fi positioning performance.

### C. FM radio

There are only few papers dedicated to FM radio based positioning. The first paper describing a localization system based on FM radio signals was presented by Krumm et al. [21]. It was an outdoors-only system that employed a prototype wristwatch device with an FM receiver. The device was able to distinguish six districts of Seattle using the signals broadcast from public FM stations. The authors were able to identify the correct district in 80% of the cases. A Bayesian algorithm with data smoothing, combined with signal propagation modeling, enabled the system to locate the user with 8 km median accuracy [4].

Fang et al. [15] presented a comparison of FM and GSM outdoor localization within 20 reference points in an urban area of 1 km2. Using the data collected with a professional spectrum analyzer, the authors demonstrated that with six-channel fingerprints the GSM accuracy was better than that of FM; however, when the number of FM channels was increased to 11 the situation reversed (FM error was below 20 m in 67% of cases). In a rural area however, GSM signals were weaker and 5-channel FM positioning outperformed the 8-channel GSM based system; the FM positioning error was within 35 m with 67% probability. The reported data does not allow for direct comparison of FM accuracy in urban and rural areas for an equal number of channels. The authors also investigated temporal stability of FM signals and reported better temporal stability of FM signal than the GSM signal [15].

More recently [16], the same group evaluated the positioning performance of multiple wireless technologies (FM, GSM, DVB, Wi-Fi) for both outdoor and indoor settings. However, FM measurements were performed only outdoors and thus FM positioning was not included into comparison to other indoor localization systems.

All the systems described so far utilize the differences of signal strength between different locations. The two main sources of signal attenuation (leading to spatial variation of fingerprints) in outdoor settings are: free-space propagation loss (in order of 20log(d), where d is travel distance) and shadowing by terrain and buildings [10]. In [15], the distance between test points was about 100 m, and free-space propagation loss contributed about 40 dB to the signal strength differences between locations. At indoor scales, however, the free-space propagation loss is negligible and the main source of spatial signal variation is fading, caused by large indoor obstacles and multipath propagation [10]. Therefore, since current FM positioning systems rely on outdoor-only propagation phenomena, their results cannot be simply extrapolated to indoor scenarios.

Another approach for FM localization utilizes the phase of the stereo pilot tone – a stable 19 kHz signal contained in all stereo FM transmissions. Giordano et al. [20] proposed an FM based outdoor localization system which leverages differences of FM stereo pilot phase, as received by the mobile unit and a fixed observer. The authors claimed the accuracy "on the order of 10–20 m depending on channel conditions" [20]. However, the origins of these numbers are questionable, since the authors have not provided an experimental proof of the claimed performance. Moreover, there are experimental indications that the pilot tone, although transmitted with a good stability, is distorted by multipath propagation [26] and non-linear effects in the receiver [13]. For instance, typical peak-to-peak pilot phase fluctuations reported in [26] were about 2 µs, which corresponds to about 600 m distance for a 19 kHz pilot tone. Such low accuracy is clearly insufficient for indoor positioning.

As the literature review shows, the previous research work on FM positioning has focused mainly on outdoor localization using broadcast FM signals and special receivers (prototype wristwatch [4] and professional spectrum analyzer [15]). This paper, in contrast, focuses on indoor positioning, using off the shelf mobile devices. There is only one work that investigated indoor positioning using FM radio signals [11]. The described system is based on a set of low-power FM transmitters installed indoors. This paper, in contrast, presents a positioning system that uses broadcasting stations located outdoors for indoor positioning. According to literature review and to the best of our knowledge, this is the first study to use broadcasting FM radio stations for indoor localization.

### III. FM POSITIONING

Our approach on FM radio based localization is based on a well-known signal fingerprinting method [9, 14]. This method includes two stages: calibration and localization. The calibration phase comprises the acquisition of signal characteristics (typically RSSI) from stationary transmitters (beacons) at predefined points, which are used to build a database that matches the collected values (fingerprints) with their corresponding locations. During the localization phase, the mobile device acquires a signal fingerprint and the positioning system utilizes the calibration data, coupled with appropriate algorithms to determine the best match for a location where the fingerprint most likely belongs.

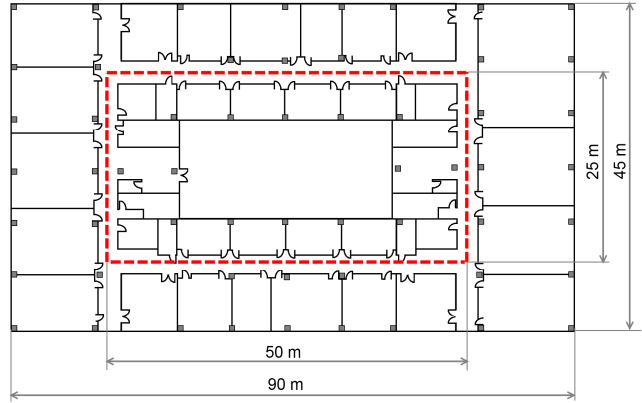

**Figure 1. Floor-level experimental environment. Dashed rectangle (50×25 m) represents the data collection path.**

For localization we used three well-known methods: k-nearest neighbor (kNN), support vector machine (SVM) classifiers, and Gaussian processes (GP) regression [9, 12, 14].

*A. Experimental setup*

The positioning experiments have been performed in two separate indoor environments of different scales: an office (12×6 m) [11] and a university building floor along its corridor (50×25 m), similar to the approach in [9]. Signal fingerprints were simultaneously acquired by three smartphones: two Samsung Omnia2 and one HTC Artemis; each device used its standard headset as antenna. Two datasets per each environment have been collected on different days. Before each data acquisition session, the list of active FM channels, broadcasting content, was automatically acquired using receiver's channel detection functionality. Only the stations simultaneously present in both training and testing dataset were considered for analysis (50 stations for room-level and 45 for the floor-level environment). At each location, 10 RSSI samples were collected for each FM channel and their mean values were used in the fingerprint. In the room-level environment, signal fingerprints have been collected in all the accessible locations defined by a 1 m grid (due to the furniture, 33 locations were measured). In the building floor environment, fingerprints were acquired with 1.6 m interval along the main corridor (dashed line in Figure 1); in total, 94 locations were measured. In both settings, the experimenter was always facing in the same direction. Localization accuracy results obtained using the same device are shown in the section that follows.

*B. Same-device localization results*

The positioning accuracy of the system was first evaluated using independent datasets collected with the same device and evaluated using kNN, SVM and GP methods. All the collected RSSI values were normalized in the range of 0 to 1. The performance results of the FM positioning system using broadcasting FM stations are presented in Figure 2. At the room level, 40% of locations were successfully recognized by the SVM classifier (0 m error distance), while the best median positioning error of the system was 0.6 m (SVM) and the 90th percentile error was 2.6 m (GP). In the floor-level environment, the best results were demonstrated by the kNN approach, which accurately recognized 52% of locations; with 90% probability, the error was below 24 m. The GP method demonstrated a poor performance, due to continuous nature of regression resulting in majority of the two-dimensional location estimates to fall *inside* the data collection rectangle rather than *on* it (see Figure 1).

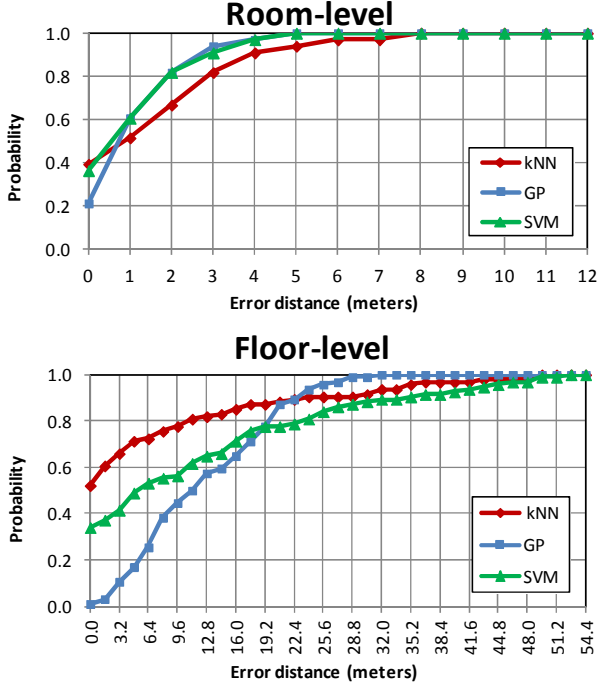

**Figure 2. FM positioning accuracy for different methods.**

While a detailed analysis of the results will follow in Section III.F, the presented performance can be preliminary attributed to the high number of FM beacons used, despite the significant distance from the transmitters. There is evidence in the literature to suggest that wider fingerprints, collected from a large number of beacons, result in better positioning performance [15, 17]. In order to test the impact of number of beacons on positioning accuracy, we carried out a set of experiments to evaluate the effect of number of stations on localization accuracy.

### C. Accuracy vs. number of stations

This section describes a number of approaches to FM station selection and evaluates how the number of stations (fingerprint width) affects the positioning accuracy.

#### 1) Heuristic-based approach

Initially, we briefly experimented with several station selection criteria, which would enable identification of the most efficient (in terms of localization performance) stations. One criterion was to select N strongest stations, while the other criterion was to select N weakest stations. However, none of these methods was able to demonstrate a consistent advantage in all environments.

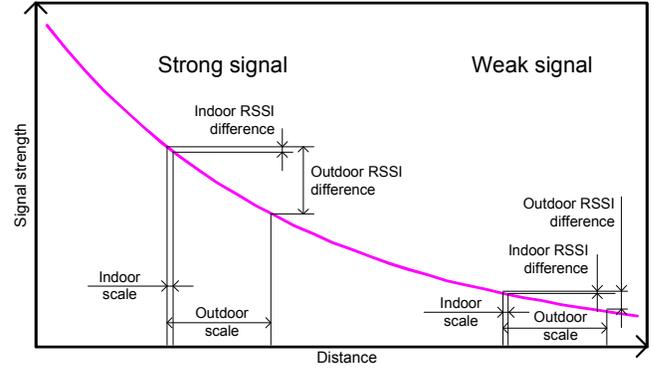

**Figure 3. RSSI difference caused by path loss at indoor and outdoor scales.**

These findings are somewhat in contradiction with the results previously demonstrated for outdoor FM positioning. Fang et al. [15] have found that in outdoor scenarios selecting stations with stronger signals leads to better positioning accuracy than for weak-signal stations. To understand the reasons for this inconsistency, we analyzed the difference between indoor and outdoor positioning using broadcasting beacons. In the outdoor scenario, the distances between test points are relatively large (100–150 m in [15]). At these distances the transmission of nearby stations with strong signal are subject to significant path loss (Figure 3).

In indoor environments, however, the distances between test points are orders of magnitude smaller (1–2 m in our tests), and the path loss has minimal effect on signal propagation (see Figure 3). In this case, the variance of FM signal RSSI in indoor locations is due to walls and other large obstacles that equally affect all stations transmitting from the same direction, regardless of their signal strengths. Therefore, in indoor scenarios stronger stations have no advantage over weaker stations in terms of positioning performance. A further analysis of this phenomenon is presented in Section III.F.

#### 2) Statistical approach

Although the previous approach did not suggest a definite criterion for station selection, it would be useful to at least assess the extent to which the number of stations influences the localization performance. In order to estimate this relationship, we used a statistical approach and evaluated FM localization performance on $N$ randomly selected non-repeating channels. For each value of $N$, 500 localization tests have been run. Fingerprint recognition was performed by the kNN method, as the computational time of other methods was prohibitive for large number of trials.

The experimental results in Figure 4 demonstrate a dependence of median error on the number of stations used for positioning. The results confirm hypothesis that localization accuracy consistently improves when increasing the number of stations used for fingerprinting. Moreover, the stations contribute to the performance unequally: for example, in the room environment, five selected stations were able to provide zero median error, in contrast to the average 1.8 m for randomly chosen stations.

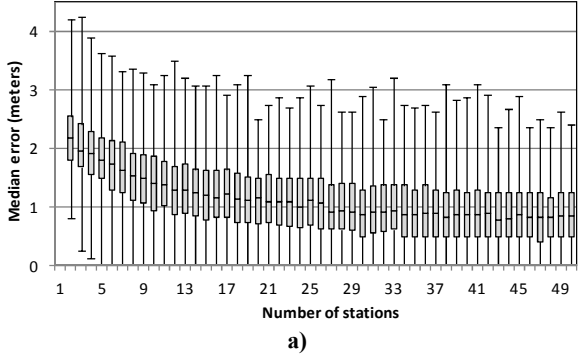

a)

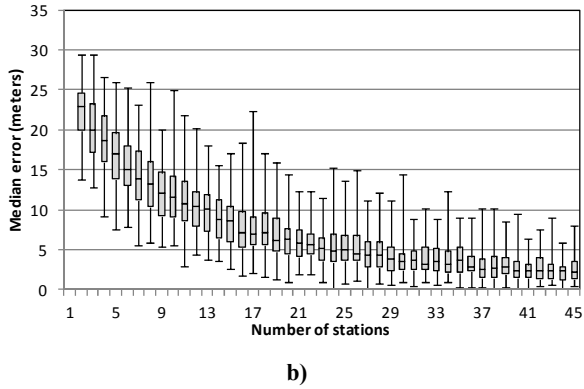

b)

**Figure 4. Median error of FM positioning vs. number of stations in room-level (a) and floor-level (b) environments.**

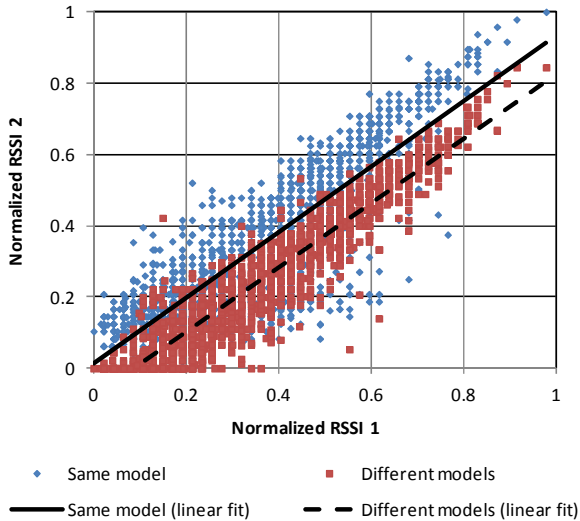

**Figure 5. Scatter plots of RSSI values measured in the building floor environment by co-located smartphones: two Samsung Omnia2 ("same model") and HTC Artemis vs. Omnia2 ("different models").**

Therefore, appropriate selection of broadcasting FM stations can reduce the fingerprint width (and acquisition time) with only minor degradation of positioning accuracy. Moreover, reducing the number of scanned stations enables faster acquisition of fingerprints and decreases the overall computational requirements of the positioning system. For our test devices, by decreasing the number of stations from 50 to 5, we reduce the acquisition time from 10 s to 0.7 s; however, this increases the average median error by only 1 m (Figure 4a). While floor-scale FM localization would typically require more stations, a careful selection of stations would still reduce fingerprint acquisition time, although to a lesser extent.

### D. Handling device diversity

All the experimental results reported so far relied on data collected with the same device. However, fingerprints collected by different devices can be considerably different due to the diversity of hardware and software designs, manufacturing processes and fluctuations of the components' parameters. Therefore, even devices of the same model, put in the same conditions, may report different RSSI values. Measurements in Figure 5 show that two devices of identical model, report different RSSI values, shown as RSSI1 and RSSI2 in the "same model" scatter plot. Clearly, having differences between fingerprints collected during the system calibration phase and fingerprints acquired for localization, may severely impact the localization accuracy.

The literature suggests several approaches to address this problem, such as estimation of a cross-device RSSI mapping function or switching to different fingerprint types [23-25]. In this paper, we compare three approaches to cross-device location sensing. The first method (labeled "basic") does not take device diversity into account and uses raw RSSI fingerprints directly (as demonstrated in [9]). The second method ("ratio") relies on hyperbolic fingerprints proposed by Kjaergaard and Munk [23]. A hyperbolic fingerprint is composed of pairwise ratios of RSSIs instead of raw values. In [23] such fingerprints were shown to improve the localization accuracy of kNN and Bayesian classifiers. Finally, the third method ("correlation") originates from the work of Tsui et al. [24], which assumes that while raw fingerprints collected by different devices might be rather distant in the signal space, they are still highly correlated (similar in shape). Instead of Euclidean distance, this method employs Pearson's correlation coefficient as a measure of similarity between fingerprints.

To evaluate cross-device performance of FM localization, we trained a kNN classifier using a dataset collected by one of the Omnia2 smartphones, and then tested it with datasets collected on a different day with HTC Artemis and another Omnia2 device ("different models" and "same model" titled graphs in Figure 6, respectively).

Figure 6 presents cross-device, over-time performance of FM localization system. At a room level, device heterogeneity plays a minor role: the same-model performance is comparable to the same-device results (Figure 6a); for different device models, the classification rate of the basic method drops by 12%, while the other performance metrics remain comparable to the same-device reference (Figure 6b). At a larger scale, the degradation of cross-device performance becomes more evident. For basic localization approach, which does not take device diversity into account, the classification rate drops from 52% (same-device) to 37% (same model, Figure 6c) and even 28% (different models, Figure 6d).

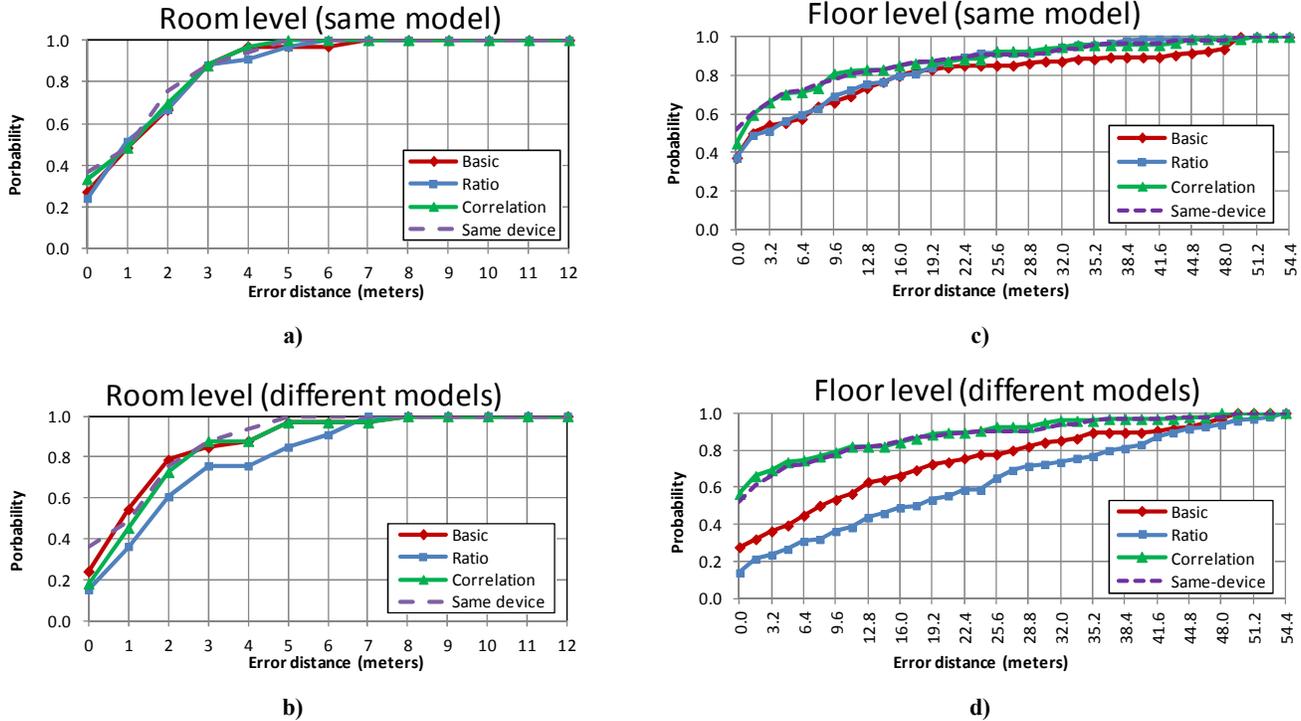

Figure 6. Performance of an FM localization system trained on data from one Omnia2 device and tested on different day with another Omnia2 ("same model" graphs) and with HTC Artemis ("different models" graphs) smartphone. Dashed line presents the reference performance: testing with the same Omnia2 device that was used for training.

The localization methods devised for handling cross-device differences demonstrated diametrically opposite results with respect to each other. The hyperbolic fingerprinting approach demonstrated performance similar to the basic approach (in same-model tests) or even lower performance (for different-model tests). The best cross-device performance was demonstrated by the "correlation" approach, which estimated fingerprint similarity using correlation coefficient instead of Euclidean distance. This method was able to achieve a cross-device FM localization performance comparable to the baseline (same-device) tests both at room and floor scales (see Figure 6).

The reasons behind different performance of the described approaches originate from how they estimate cross-device RSSI mapping. The ratio method takes into account only the coefficient parameter of the linear RSSI mapping between devices, but not the bias [23, 24]. However, a non-zero RSSI mapping offset between different models of experimental devices is present and can be seen in Figure 5. Moreover, the ratio method boosts fingerprint width by including all pair-wise RSSI ratios; thus, eventual mapping errors and outliers are duplicated a number of times, which decreases the classification accuracy below the basic performance reference. In contrast, Pearson's correlation coefficient implicitly considers both parameters of a linear fit and is less sensitive to eventual outliers, explaining the high performance of the correlation based approach.

### E. Comparison to Wi-Fi and GSM

This section presents a comparison of FM localization accuracy with Wi-Fi and GSM based systems. Wi-Fi is the de-facto standard of indoor localization, while GSM and FM share similar infrastructure characteristics with respect to the stations being outside of the test environment.

Due to the factors influencing the performance of a positioning system, such as room layout and hardware characteristics, it is very difficult to directly compare the positioning performance results acquired in different environments. Therefore, to ensure a fair comparison, Wi-Fi, GSM and FM RSSI fingerprints were collected simultaneously, in the same environment, during the experiments described in the previous sections. For each -location, 10 Wi-Fi RSSI samples were acquired with 1 s interval. In total there were 15 different Wi-Fi beacons in the room-level dataset and 65 in the floor-level dataset. Average fingerprint widths per location were 10 and 21, respectively; many of Wi-Fi beacons had very low signal levels and were visible only in few locations. GSM RSSI fingerprints for 7 nearby GSM base stations were collected only by the HTC Artemis smartphone, because Omnia2 devices do not provide access to GSM module. For each test point, 3 GSM samples were recorded with 5s interval, which was the maximum update rate of device's GSM information. This resulted in total of 30 different cell IDs in the room dataset and 46 cell IDs in the floor-level dataset; average fingerprint widths were 8 and 7 beacons per location, respectively. As described before, the FM fingerprints contained 10 RSSI samples from 50 FM stations for the room level and 45 FM stations for the floor level. All the collected RSSI values were normalized in 0…1 range and consequently the fingerprints at each location were averaged to obtain a single fingerprint for each wireless technology; missing RSSI readings were ignored in the averaging procedure. RSSI values for the beacons that were not visible at a location were set to zero. Localization performance has been evaluated with the kNN approach using same-device different-day datasets.

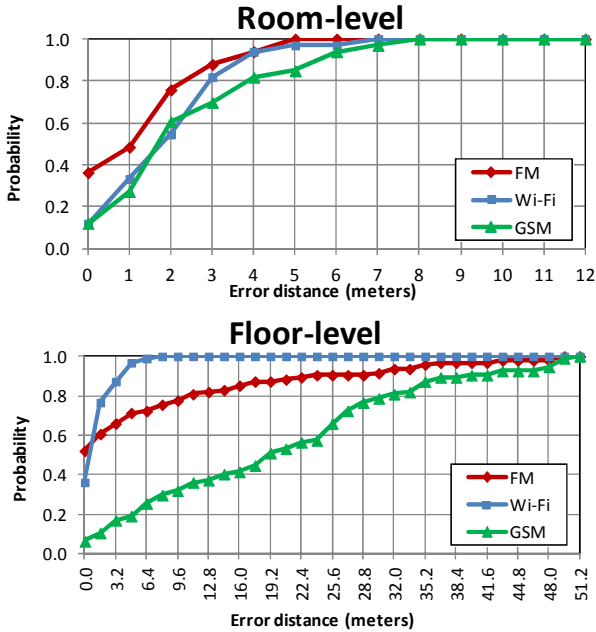

**Figure 7. Comparison of FM, Wi-Fi and GSM positioning accuracy.**

The results are presented in Figure 7. In the room environment, FM demonstrated the lowest median error (0.9 m) in comparison to other localization technologies. At higher probability levels, FM performance was comparable to Wi-Fi (4.0 m at 95% probability). For the GSM approach, median error was comparable to Wi-Fi, however at higher probabilities GSM accuracy was lower in comparison to other technologies.

At larger scale environment, the ranking of the positioning systems notably changes. While Wi-Fi accuracy remains almost the same (3.6 m at 95% probability), the performance of the localization technologies based on external infrastructure drops significantly. Although the FM-based system was able to correctly recognize 52% of locations, its 95[th] percentile error increased to 34.6 m. The GSM approach, in turn, showed even higher degradation of performance (48 m at 95[th] percentile). GSM results are consistent with a previous report about 6-strongest-cell GSM indoor localization [17].

Therefore, the above results demonstrate that while Wi-Fi fingerprinting exhibits stable performance in both settings,

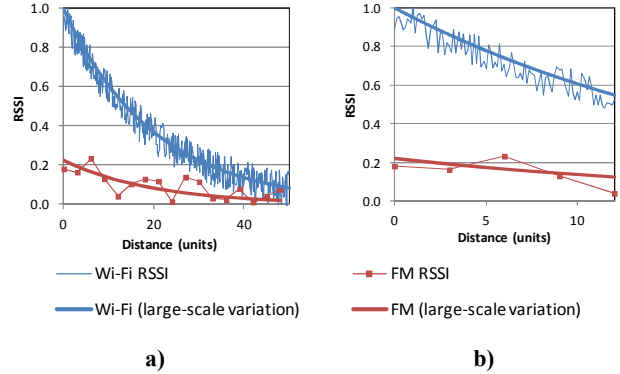

**Figure 8. An example of spatial RSSI distribution for Wi-Fi and FM radio waves in a large-scale (a) and small-scale (b) indoor environment.**

FM localization provides higher accuracy in small confined environments, while at a larger scale the performance drops. In the section that follows we analyze possible reasons of such behavior.

*F. Analysis of results*

The signal fingerprinting approach relies on the fact that radio signals vary depending on location. This dependency, however, includes several components, main of which are large-scale and small-scale variations [5,18]. *Large-scale variations* are caused by the fact that the signal intensity is inversely proportional to the square of transmitter-to-receiver distance (free-space path loss) [10, p. 11]. *Small-scale variations*, in turn, happen due to multipath propagation of the radio signals caused by radio wave interactions (reflection, diffraction, interference) with indoor objects [5]. Small-scale variations occur at distances comparable to wave length (~0.12 m for Wi-Fi, ~3 m for FM).

Now, we consider the role of small-scale and large-scale variations in FM and Wi-Fi based indoor localization.

Wi-Fi access points are typically installed within the indoor environment (building) where they are used; they have a limited transmission power and limited coverage. Therefore, Wi-Fi RSSI can noticeably change within few meters due only to free-space path loss [18]. In turn, small-scale variations of Wi-Fi signals (~12 cm wavelength) can be considered as a noise superimposed on the large-scale variations (see Figure 8a). Thus, in both experimental environments Wi-Fi localization relied mainly on large-scale variations and therefore demonstrated similar performance in room and floor scale.

FM radio stations, in contrast, have significantly higher transmission power and coverage. They are typically located at a considerable distance from listeners, so that received signals correspond to the almost flat tail of the free-space path loss curve (see Figure 8a). As a result, even substantial changes in distance vary the free-space path loss only slightly; so, large-scale variations of ambient FM radio signals are minimal at indoor distances of several meters. In contrast, small-scale variations play a significant role in small indoor areas, such as rooms, due to the large wave length of FM radio waves (~3 m). In a small confined area, the small-scale variations create unique enough signal distribution patterns, which results in high accuracy of the fingerprint-based localization (Figure 8b). In larger environments, however, the patterns of small-scale variations are ambiguous (Figure 8a) and the localization performance decreases.

In terms of signal variation, GSM base stations are similar to FM stations: they are situated at a considerable distance from the mobile devices, and have a medium coverage. Thus, large-scale variations of GSM signals are likely to be more expressed than for FM, but less than for Wi-Fi. The small-scale variations of GSM signals have a scale of 0.3 m (for 900 MHz GSM band), and signal patterns might become ambiguous even in a rather small area. On the other hand, a regular GSM phone reports RSSI of up to 7 strongest nearby stations, while FM fingerprints may contain dozens of channels and thus provide better localization accuracy (see Section III.C).

## IV. POWER CONSUMPTION

In typical ubiquitous computing applications, available battery power is a major limitation that determines sampling rate, data collection length and general usability of an application. Therefore, modules with low power requirements not only increase the amount of data that can be collected, but also allow long-term monitoring, while having a minimal impact on user's device. Considering the importance of power consumption, we measured the battery life of a mobile device in FM and Wi-Fi localization modes.

The experiments were carried out on a Samsung Omnia2 device. During the tests, the phone periodically acquired location fingerprint and stored it in a database. It was not possible to acquire GSM RSSI because the device did not provide an access to the phone module. However, the base stations are automatically queried by the GSM module for its internal operation and no special software-initiated scanning is required (in contrast to FM and Wi-Fi). Therefore, battery life in GSM mode was estimated by leaving the device with enabled GSM module at a place with a medium signal level. During the tests, the device was kept in an "unattended" power state [27], with screen and unused wireless modules turned off. The tests started with a fully charged battery and continued until the device run out of power and turned itself off. The baseline battery performance was acquired with all the wireless modules switched off.

Figure 9 presents the results of battery life measurements in FM and Wi-Fi fingerprinting modes for different time intervals between consecutive fingerprints. It was not possible to evaluate Wi-Fi performance with 1 s update period, as the Wi-Fi driver became unstable at high sampling rates and often crashed. From the general trend, however, it is evident that the result would have been less than 7.4 h, which is still significantly below the 27.9 h demonstrated by FM with 3 beacons. With 10 or more seconds between scans, the FM reaches its maximum performance, providing only 1.3 h (3%) shorter battery life in comparison to the baseline. In contrast to FM, Wi-Fi puts a significant load on the battery, and demonstrates only 7.4 h and 12.6 h battery life with 10 s and 20 s interval, respectively.

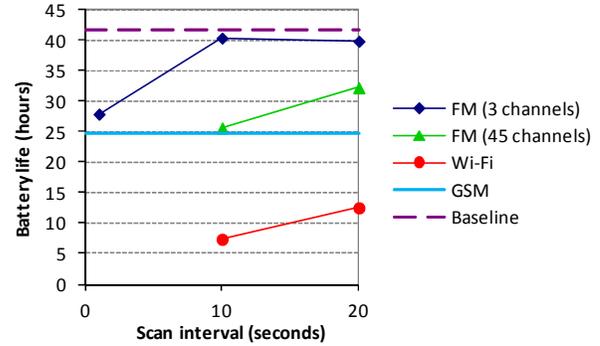

**Figure 9. Battery life in FM and Wi-Fi fingerprinting modes.**

Overall, FM demonstrates superior power efficiency, providing 2.6 to 5.5 times longer battery life than Wi-Fi and closely approaching the baseline maximum.

## V. CONCLUSION

This paper presented an investigation and evaluation of FM broadcasting stations as an alternative for indoor localization. The experimental results shown have demonstrated the feasibility of indoor localization using broadcasting FM radio stations. We have provided an experimental evaluation of limitations of external infrastructure (such as FM or GSM) when the localization area increases, and complemented it with an analysis of the location fingerprints dependence on signal wave lengths. We have also experimentally verified that FM receivers have significantly lower power consumption than Wi-Fi modules and provide 2.6 to 5.5 times longer battery life in localization mode. While FM radio technology in our experiments required headphones to serve as an antenna, this is not a fundamental limitation since certain devices (such as portable MP3 players) are typically used with headphones; moreover, FM tuner chips are sensitive enough to be used with embedded antennas [28]; and finally, there are already devices on the market with internal FM antennas (for example, Nokia 5030).

This work is the first to investigate indoor localization using FM broadcasting stations and thus a number of future research directions remain to be followed, including the influence of user facing direction, device holding position or antenna configuration on RSSI readings.

## REFERENCES

[1] A.K. Dey and G.D. Abowd, "Towards a better understanding of context and context-awareness," Proc. CHI 2000, ACM Press, 2000.

[2] H. Liu, H. Darabi, P. Banerjee, and J. Liu, "Survey of wireless Indoor positioning techniques and systems," IEEE Trans. Systems, Man, and Cybernetics, Part C. 37, 2007, pp. 1067-1080.

[3] "Interference mitigation challenges and solutions in the 2.4 to 2.5-GHz ISM band". http://www.cypress.com/?docID=4665.

[4] A. Youssef, J. Krumm, E. Miller, G. Cermak, and E. Horvitz, "Computing location from ambient FM radio signals," Proc. WCNC 2005, IEEE, 2005, pp. 824-829.

[5] H. Hashemi, "The indoor radio propagation channel," *Proc. IEEE*, vol. 81, 1993, pp. 943-968.

[6] International Telecommunication Union, "Recommendation ITU-R P.840-4," 2009.

[7] International Telecommunication Union, "Recommendation ITU-R P.833-6," 2007.
[8] International Telecommunication Union, "Recommendation ITU-R P.620," 2005.
[9] P. Bahl and V. Padmanabhan, "RADAR: An in-building RF-based user location and tracking system," Proc. INFOCOM 2000, IEEE, 2000, pp. 775-784.
[10] L.W. Barclay, "Propagation of radiowaves," IET, 2002.
[11] A. Matic, A. Popleteev, V. Osmani, and O. Mayora-Ibarra, "FM radio for indoor localisation with spontaneous recalibration," Pervasive and Mobile Computing, Elsevier, vol. 6, 2010, pp. 642–656.
[12] B. Ferris, D. Haehnel, and D. Fox., "Gaussian processes for signal strength-based location estimation," Proc. RSS 2006, MIT Press, 2006.
[13] US Federal Communications Commission, "Docket No.13506: FM stereo final report and order," 1961.
[14] M. Brunato and R. Battiti, "Statistical learning theory for location fingerprinting in wireless LANs," Computer Networks, Elsevier, vol. 47, 2005, pp. 825-845.
[15] S.-H. Fang, J.-C. Chen, H.-R. Huang, and T.-N. Lin, "Is FM a RF-based positioning solution in a metropolitan-scale environment? A probabilistic approach with radio measurements analysis," IEEE Trans. Broadcasting, vol. 55, 2009, pp. 577-588.
[16] S. Fang and T. Lin, "Cooperative multi-radio localization in heterogeneous wireless networks," IEEE Trans. Wireless Communications, vol. 9, 2010, pp. 1547-1551.
[17] V. Otsason, A. Varshavsky, A. LaMarca, and E. de Lara, "Accurate GSM indoor localization," Proc. UbiComp 2005, Springer, 2005, pp. 141–158.
[18] M. Youssef and A. Agrawala, "Small-scale compensation for WLAN location determination systems," Proc. WCNC 2003, IEEE, 2003, pp. 1974–1978.
[19] A. Varshavsky, E. de Lara, J. Hightower, A. LaMarca, and V. Otsason, "GSM indoor localization," Pervasive and Mobile Computing, Elsevier, vol. 3, 2007, pp. 698–720.
[20] A. Giordano, D. Borkowski, and D. Kelley, "Location enhanced cellular information services," Proc. PIMRC 1994, IEEE, 1994, pp. 1143-1145.
[21] J. Krumm, G. Cermak, and E. Horvitz, "RightSPOT: A novel sense of location for a smart personal object," Proc. UbiComp 2003, Springer, 2003, pp. 36-43.
[22] A. Varshavsky, M.Y. Chen, E. de Lara, J. Froehlich, D. Haehnel, J. Hightower, A. LaMarca, F. Potter, T. Sohn, K. Tang, and I. Smith, "Are GSM phones THE solution for localization?" Proc. IEEE Workshop on Mobile Computing Systems and Applications (WMCSA), IEEE Computer Society, 2006, pp. 20–28.
[23] M.B. Kjærgaard, and C.V. Munk, "Hyperbolic location fingerprinting: A calibration-free solution for handling differences in signal strength," Proc. PerCom 2008, IEEE, 2008, pp. 110-116.
[24] A.W. Tsui, Y.-H. Chuang, and H.-H. Chu, "Unsupervised learning for solving RSS hardware variance problem in WiFi localization," Mobile Networks and Applications, vol. 14, 2009, pp. 677-691.
[25] A. Haeberlen, E. Flannery, A. Ladd, A. Rudys, D. Wallach, and L. Kavraki, "Practical robust localization over large-scale 802.11 wireless networks," Proc. MobiCom 2004, ACM, 2004, pp. 70-84.
[26] D. Howe, "Precise frequency dissemination using the 19-kHz pilot tone on stereo FM radio stations," IEEE Trans. Broadcasting, vol. BC-20, 1974, pp. 17-20.
[27] Microsoft Corp., "Windows Mobile 6.5 system power states". http://msdn.microsoft.com/en-us/library/aa930499.aspx.
[28] Silicon Laboratories, Inc. "Si4704/05 datasheet," 2009.